
%
\baselineskip 9truemm
\vsize 21.0 truecm
\hsize 14.0 truecm
\overfullrule=0truept
\def\ltsim{\raise 2pt \hbox {$<$} \kern-1.1em \lower 4pt \hbox {$\sim$}}
\def\ltapprox{\raise 2pt \hbox {$<$} \kern-1.1em \lower 5pt \hbox {$\approx$}}
\def\gtsim{\raise 2pt \hbox {$>$} \kern-1.1em \lower 4pt \hbox {$\sim$}}
\def\gtapprox{\raise 2pt \hbox {$>$} \kern-1.1em \lower 5pt \hbox {$\approx$}}

\def\skdue{\vskip 50pt}
\skdue
\centerline {\bf VLBI Observations of a complete sample of radiogalaxies.}
\centerline {\bf VI-The Two FR-I Radio Galaxies B2 0836+29 and 3C465}
\skdue
\centerline{T.Venturi$^{1}$, C. Castaldini$^{1}$, W.D. Cotton$^{3}$
L.Feretti$^{1,2}$, G.Giovannini$^{1,2}$,}
\par\noindent
\centerline{L.Lara$^{1,4}$, J.M. Marcaide$^{5}$, A.E.Wehrle$^{6}$}
\skdue
\noindent 1: Istituto di Radioastronomia  del CNR
\par\noindent \ \ \ Via Gobetti 101, I-40129 Bologna (Italy)
\par
\noindent 2: Dipartimento di Astronomia dell'Universit\'a, Bologna
\par\noindent \ \ \ Via Zamboni 33, I-40133 Bologna (Italy)
\par
\noindent 3: National Radio Astronomy Observatory,
\par\noindent \ \ \ 520, Edgemont Road, VA 22903-2475 Charlottesville, USA
\par\noindent 4: Instituto de Astrofisica de Andalucia, CSIC,
\par\noindent \ \ \ Apdo. 3004, 18080 Granada, Spain
\par\noindent 5: Departamento de Astronomia, Universidad de Valencia
\par\noindent \ \ \ 46100 Burjassot, Spain
\par\noindent 6: Infrared Processing and Analysis Center, Jet Propulsion
Laboratory
\par\noindent \ \ \ California Institute of Technology, M/S 100-22 CA91125
Pasadena, USA
\medskip
\noindent
Address for offprint request: Tiziana Venturi (Istituto di Radioastronomia)
\vfill
\eject
\centerline{\bf Abstract}
\medskip
\noindent
We present 5 GHz global VLBI observations of the two
Fanaroff Riley Type I radio galaxies
B2 0836+29 and 3C465 (2335+26). For 3C465 we present also 1.7 GHz and
8.4 GHz global VLBI data. In addition VLA observations were used to obtain
arsecond resolution continuum and polarization maps at 5 GHz.
\par\noindent
Both sources are very asymmetric on the parsec-scale, with a core and a
one-sided jet, aligned with the main arcsecond scale jet.
We place a limit on the milliarcsecond jet to counterjet brightness ratio
B$_{jet}$/B$_{cjet}$ \gtsim 20 and \gtsim 30 for B2 0836+29 and
3C465 respectively.
For 3C465 the strong asymmetry holds to the kiloparsec scale.
\par\noindent
The brightness asymmetry and the ratio between the core radio power and
total radio power allow us to constrain the jet velocity close to the core and
the orientation of the radio structure with respect to the line of sight.
The results suggest that the plasma speed is relativistic on the parsec scale
for both sources, i.e. v$_{jet}$ \gtsim 0.75c for B2 0836+29 and v$_{jet}$
\gtsim 0.6c for 3C465. While v$_{jet}$ decreases from the parsec to the
kiloparsec scale in B2 0836+29, in 3C465 the very high v$_{jet}$ holds all
the way to the kiloparsec-scale {\t bright spot}.
\par\noindent
Our results are in agreement with the unification scheme suggestion that
FR-I radio galaxies are the unbeamed poulation of BL-Lac objects.
Furthermore, they reinforce the idea that the central engine in
FR-I and FR-II radio galaxies must be
qualitatively similar. The different radio morphology could then be due either
to an intrinsically  different nuclear power, which affects the torus geometry
or to different conditions in the region beyond the parsec scale, where a
significant deceleration in the FR-I jets occurs.
\medskip
\noindent Subject headings: galaxies: individual (B2 0836+29, 3C465) -
galaxies: nuclei - radio - continuum: galaxies - techniques: interferometric
\vfill
\eject
\noindent
{\bf 1. Introduction}
\medskip
\noindent
This is the sixth of a series of papers, whose main aim is to present and
discuss the observational and physical properties of the parsec-scale
radio emission in a sample of low-intermediate luminosity
radio galaxies.
\par\noindent
VLBI observations of radio galaxies are critical
to address a number of questions, such as the properties of AGNs and
the validity of the unified schemes. According to the unified
models, the low luminosity Fanaroff-Riley Type I (FR-I) radio galaxies
(Fanaroff \& Riley, 1974) are expected to be the
parent population of BL-Lacertae type objects,
while the more powerful FR-II radio galaxies would be the parent population
of radio loud quasars, the difference in their
observed properties being entirely due to orientation effects (see Antonucci,
1993 and references therein, for an updated  review of the current
models and ideas). On the other hand, the intrinsic properties, such as
for example the
Lorentz factor $\gamma$, should be the same in the two classes.
\par\noindent
To investigate these topics we are observing a sample of radio galaxies
at VLBI resolution. The sample has been presented in Giovannini, Feretti
\& Comoretto (1990, Paper I), while a detailed discussion on individual sources
observed thus far has been presented in the following papers:
NGC315, Venturi et al., 1993, Paper II; 3C338, Feretti et al., 1993,
Paper III; NGC2484, 3C109 and 3C382, Giovannini et al., 1994, Paper IV;
4C31.04 and 3C346, Cotton et al., 1995, Paper V.
\par\noindent
In this paper we present VLBI and VLA observations of two FR-I radio
galaxies in the sample, B2 0836+29 and 3C465, associated with the brightest
cluster member
at the center of the rich Abell cluster of galaxies A690 and A2634
respectively.
\par\noindent
While FR-I radio galaxies are well studied on the large
(kiloparsec) scale, little is known about their parsec-scale
properties, mainly as a consequence
of their low power radio core, which has severely limited radio
observations. According to current theoretical models by
Laing (1993) and Bicknell (1995), jets in FR-I radio galaxies are expected to
be relativistic
at their origin, as in FR-IIs, however they are strongly decelerated on the
kiloparsec scale. With the present paper we want to add further information to
the parsec-scale plasma speed in low luminosity radio galaxies.
\par\noindent
A Hubble constant H$_0$= 100 km sec$^{-1}$Mpc$^{-1}$ and a deceleration
parameter q$_0$ = 1 are assumed throughout the paper.
\vskip 1.0truecm
\noindent
{\bf 2. The Sources}
\medskip
The two radio galaxies investigated here have very similar
properties.
They are both identified with the dominant member of a rich Abell cluster,
and their total radio power is close to the upper limit of the FR-I radio
power range.
Their large scale radio structure shows distortions,
typical of {\it wide angle tail} radio sources (Owen \& Rudnick, 1976),
probably due to the interaction of the hydrodynamic flow
with the surrounding cluster medium
(O'Donoghue, Eilek \& Owen, 1993).
This class of radio sources is
characterised by radio power and morphological properties typical both of
FR-Is and FR-IIs.
The present sources have a rather bright core and two-sided jets,
strongly asymmetric in brightness, with the main jet well collimated
and with a small opening angle, as it is typical for
jets in FR-II radio galaxies. In both cases one of the two jets ends in
one {\it bright spot}, similar to the hot spots which
characterise FR-II radio galaxies. Both sources are characterised by
diffuse and distorted lobes, as it is observed in FR-Is
\medskip
\noindent
{\bf 2.1 B2 0836+29 }
\par\noindent
B2 0836+29  (4C29.30) is an extended, {\it wide angle tail}
radio source. Its optical counterpart
(RA$_{(1950)}$ = $08^h36^m13.571^s$, DEC$_{(1950)}$ =
$29^{\circ}01^{\prime}14.76^{\prime \prime}$)
is a multiple nuclei cD galaxy (Owen, Rudnick \& Peterson, 1977),
at a redshift of 0.0788,
dominating the rich cluster of
galaxies Abell 690 (Owen, White \& Thronson, 1988). The
absolute optical magnitude of the galaxy is M$_V$ = -22.4
(Giovannini et al., 1988).
\par\noindent
The large scale radio properties of B2 0836+29 were studied in detail
by many authors (Fanti et al., 1987, O'Donoghue, Owen \& Eilek, 1990,
Capetti, Fanti \& Parma, 1995, hereafter CFP95).
Its total radio power at 408 MHz is $1.2 \times 10^{25}$ W/Hz (logP = 25.08).
The kiloparsec-scale radio emission is dominated by two lobes: the southern
lobe
has a tail-like shape extended $\sim 4^{\prime}$, ($\sim$ 250 kpc), while the
northern one extends for $\sim 5^{\prime}$ ($\sim$ 300 kpc) and shows only
minor distortions.
The radio galaxy jets are asymmetric on this scale, with
the counterjet shorter and weaker than the main jet. The main jet is not
straight, but shows oscillations of few degrees amplitude around an average
position angle of $\sim 0^{\circ}$. The jet properties
and the jet/counterjet brightness ratio were discussed in
CFP95, who also presented and discussed
the polarization properties. The plasma speed along the jet is estimated
to be relativistic out to a distance of $\sim$ 5 kpc, then entrainment
from the external medium causes deceleration.
The radio emission in the source is polarized both along the jet and in the
low brightness tails. The magnetic field geometry along the main jet
switches from longitudinal to transverse, to longitudinal again.
No radio variability in the radio galaxy core has been reported.
\par\noindent
The radio galaxy is embedded in the intracluster gas, whose presence
is witnessed by detected extended X-ray emission.
The northern jet is coincident with
a high X-ray brightness region of the cluster,
while both tails are at the edge of the X-ray emission. Morganti et al. (1988)
report that the northern jet is overpressured, while both lobes are thermally
confined by the  intracluster gas. The nuclear X-ray emission is very faint
(Bloom and Marscher, 1991), i.e. 0.03 $\mu$Jy at 2 Kev.
\medskip
\noindent
For this galaxy we present 5 GHz global Very Long Baseline Interferometry
(VLBI) data and Very Large Array (VLA) observations simultaneous with
VLBI.
At the distance of this source, 1 mas corresponds to 1.0 pc.
\medskip
\noindent
{\bf 2.2 3C465}
\par\noindent
3C465 is associated with the giant D galaxy NGC7720 (RA$_{(1950)}$ =
23$^h$35$^m$58.1$^s$,
DEC$_{(1950)}$ = 26$^{\circ}$45$^{\prime}$15$^{\prime \prime}$, Dressler, 1980)
located at the center of the rich cluster of galaxies A2634
(Zhao, Burns \& Owen, 1989) at a redshift z=0.0322. The optical
associated galaxy, with M$_{V}$= -22.2  (Giovannini et al., 1988), exhibits
distorted isophotes, with two gravitationally
bound nuclei surrounded by a common envelope. De Robertis and Yee (1990)
found a weak, high ionisation emission line spectrum in the nuclear region of
the galaxy. Weak broad H$_{\alpha}$ is also present, and this strongly
suggests the presence of faint activity in the optical nucleus in 3C465.
\par\noindent
3C465 is the prototype of {\it wide angle tail} radio galaxies
(Eilek et al. 1984):
two aligned, strongly asymmetric jets are ejected from the radio
nucleus, associated with the strongest optical component within the envelope.
The fainter south-eastern jet flares out into a bright spot at
$\sim 30^{\prime\prime}$ from the core, beyond which it forms a distorted
lobe, extended  $\sim 5^{\prime}$. The main, north-western jet has a
brightness peak at $\sim 30^{\prime\prime}$ from the core, then it starts
bending and forms an extended ($\sim 5^{\prime}$), distorted lobe.
Its total power at 408 MHz is $2.0 \times 10^{25}$ W/Hz (logP = 25.30).
The radio jet, the spots and the low brightness tails are strongly
polarized. The magnetic field remains longitudinal along the whole length
of the jet, before it forms the bright spot.
\par\noindent
The core flux density of 3C465 was monitored from 1976 to 1980
(Ekers et al. 1983).
A marginal evidence ($\sim 3\sigma$ level) of nuclear variability was found.
\par\noindent
The whole radio galaxy is embedded in the X-ray emission coming from the
cluster (Burns et al., 1994).
\medskip
\noindent
We present VLBI observations of this radio galaxy at 1.7 GHz, 5 GHz
and 8.4 GHz and a VLA observation at 5 GHz, obtained during the 5 GHz VLBI
observation.
\par\noindent
At the distance of the cluster, 1 mas corresponds to 0.44 pc.
\vskip 1.0truecm
\noindent
{\bf 3. Observations, Data Reduction and Results}
\medskip
\noindent
{\bf 3.1 B2 0836+29}
\medskip
\noindent
{\it 3.1.1 VLBI and VLA Observations and Data Reduction}
\par\noindent
The source was observed in November 1990, at 5 GHz, with a
VLBI array consisting
of 8 radio telescopes. In Table 1 we report frequency, recording
system, array, u-v range, duration of the
observation and observing epoch.
The data were recorded with the MKIII recording system (Alef, 1989),
with a 28 MHz bandwidth (mode B).
and were correlated in Bonn. The final u-v coverage (after editing to
remove the bad data) is shown in Fig. 1a.
\par\noindent
The VLBI data reduction was carried out following the standard procedure.
The visibility amplitudes were calibrated
using the gain values and the system temperatures measured during the
observations at each site. Further corrections to the antenna gains,
of the order of a few percent, were
applied comparing the flux density of the VLBI calibrator as measured with
the Bonn antenna during the VLBI observations.
The calibrated data were global fringe fitted (Schwab and Cotton, 1983),
self calibrated and mapped
using the NRAO Astronomical Image Processing System (AIPS).
The final VLBI map is given in Fig. 2, and
the map parameters are given in Table 2.
\medskip
\noindent
The phased VLA took part in the VLBI observations in the
C configuration. The VLA antenna gains and phases were calibrated by means
of standard VLA calibrators.
Total and polarized VLA intensity maps were produced. Map parameters are given
in Table 3.
\medskip
\noindent
{\it 3.1.2 The parsec-scale properties}
\par\noindent
The parsec-scale morphology of B2 0836+29  (Fig. 2) consists of a one-sided
structure, i.e.
an unresolved dominant component with an elongated jet, oriented to the north
and aligned within a few degrees with the strongest kiloparsec-scale jet.
\par\noindent
The total VLBI flux is the same as the arcsecond core flux, measured by us
at the epoch of the VLBI observations, so we are confident we have mapped
the whole parsec-scale structure.
The strongest VLBI component contains
$\sim 88$\% of the total VLBI flux. Such percentage should be considered an
upper limit, since the beam is very elongated along
the jet direction and the VLBI core flux could be contaminated by the beginning
of the northern jet.
The arcsecond core spectrum is inverted between 1.4 GHz and 5 GHz
(CFP95), therefore we treat the strongest VLBI component as the core of
the radio galaxy.
\par\noindent
The radio emission along the jet shows three peaks
respectively at $\sim$ 6 mas, at $\sim$ 12 mas and at $\sim$ 16 mas from
the core.
The latter one is a well separated blob,
misaligned by a few degrees with the main jet direction.
No counterjet is visible either in the full resolution VLBI map shown here,
or in a low resolution map obtained by omitting the longest baselines.
We point out that the shape of the beam, very elongated in the north-south
direction, would hide a short counterjet, if present.
We quantify the jet/counterjet brightness asymmetry by placing a lower limit
$R = {B_{jet} \over B_{cjet}}$ \gtsim 20 at 6 mas from the core.
\medskip
\noindent
{\it 3.1.3 The kiloparsec-scale properties}
\par\noindent
In Fig. 3 we present the map obtained with the VLA at the time of the VLBI
observations. A more detailed map of the core and jets with
polarized B vectors superimposed is given in Fig. 4.
The different regimes in the northern
jet, where the polarization vector flips from perpendicular
to parallel, to perpendicular again, are clearly visible in this map.
The brightness decrease along the jet and the expansion of
the jet are in very good agreement with the results presented and
discussed in CFP95.
\par\noindent
We used our VLA map (Figs. 3 and 4) to determine the jet/counterjet
brightness ratio $R$ at different
distances from the core. We find a value  $R \sim$ 7 up to 15 kpc from the
core, $R \sim 4$ in the range 15-20 kpc, and $R \sim 2$ in the range 20 - 31
kpc (see Table 4).
Such decrease of the ratio $R$ as a function of the distance from the
core is consistent with the trend given in CFP95.
\bigskip
\noindent
{\bf 3.2 3C465}
\medskip
\noindent
{\it 3.2.1 VLBI and VLA Observations and Data Reduction}
\par\noindent
This source was observed with VLBI at three different frequencies,
i.e. 1.7 GHz, 5 GHz and 8.4 GHz, with different
recording systems. The observational parameters
are given in Table 1 and the u-v coverage is given in Figs. 1b-d.
\par\noindent
The 1.7 GHz and 8.4 GHz observations were carried out
with the MKII recording system. The telescopes taking part to the
observations are gven in Table 1.
The data were correlated with the MKII Block
2 correlator at Caltech.
The data were first global fringe fitted by means of the
task FRING in AIPS (Schwab \& Cotton, 1983), then the rest of the data
reduction was carried out
with the VLBI Caltech package (Pearson, 1991). The observed
visibilities were first
modelfitted, and the best model obtained from modelfitting was used as
starting point in the self-calibration procedure. Phase self-calibration
was applied at the beginning, and only when the phase corrections were stable
we allowed for few cycles of amplitude and phase self calibration.
Gain corrections were of the order of 5\% for both datasets.
The final maps were done in AIPS.
\par\noindent
Global VLBI observations (EVN+USVN)
at 5 GHz were carried out with the MKIII Mode B recording system
(see {\it Sect. 3.1.1}) and were correlated in Bonn. The data reduction
(global fringe fitting, calibration, editing and self-calibration)
was carried out entirely with the AIPS package.
Self-calibration started with a point source, and the final convergence
between the source model and the visibilities was reached after 12
self-calibration cycles. Amplitude self-calibration was applied only in the
last two cycles, when the phase corrections were stable. Gain corrections were
within 5\%.
\par\noindent
The contour plots of the VLBI maps at each frequency are
given in Figs. 5a-c. The 8.4 GHz map shown in Fig. 5c is convolved with a
beam larger than that of the full resolution in order to emphasise the
structure of the jet.
The observational parameters in each map are given in Table 2.
\medskip
\noindent
The VLA phased array in the C configuration was used for our 5 GHz VLBI
observations, allowing us to map the source at arcsecond scale.
The antenna gains and phases
were calibrated by means of standard VLA calibrators.
Total and polarized intensity maps are shown in Fig. 6 and 7, respectively.
\medskip
\noindent
{\it 3.2.2 The parsec-scale morphology and spectrum}
\par\noindent
The parsec-scale morphology of 3C465 is one-sided at each frequency
(Figs. 5a-c).
The parameters of the VLBI maps at the various frequencies are given in
Table 2.
The correlated flux in the parsec-scale structure is $\sim$ 95\% of
the arcsecond core flux density measured from the VLA map obtained during
the VLBI observations.
\par\noindent
The milliarcsecond-scale jet in 3C465 is on the same side as the main
arcsecond-scale jet. There is however a minor misalignment between the two,
the VLBI jet being oriented in position
angle $\sim -49^{\circ} \pm 2^{\circ}$ at the three frequencies, while
the main arcsececond-scale jet is oriented in p.a.
$\sim -56^{\circ} \pm 2^{\circ}$.
\par\noindent
The brightness along the jet decreases smoothly at 1.7 GHz, with a knot
of radio emission at $\sim 50$ mas from the peak (Fig. 5a).
In the 5 GHz and 8.4 GHz maps the jet is resolved in knots of
radio emission (Figs. 5b and 5c).
At 8.4 GHz the transverse size of the jet is at the limit of the
resolution along all its length, i.e. out to $\sim$ 4 pc from the
core. The brightness decreases from 20 mJy/beam  at $\sim$ 1 pc to
2 mJy/beam at $\sim 6$ pc.
\par\noindent
We derived the
jet/counterjet brightness ratio from the 8.4 GHz map, since the high resolution
of the map, coupled with the knotty structure of the source at this frequency,
enables us to constrain the ratio with better accuracy. At $\sim$ 1 pc
from the peak ($\sim 2.3$ mas), we obtain a lower limit
$R = {B_{jet} \over B_{cjet}}$ \gtsim 30.
\medskip
\noindent
The integrated spectrum of the VLBI structure, derived from our full
resolution total intensity maps ( see Table 2), is peaked around 5 GHz, being
$\alpha^{5 GHz}_{1.7 GHz} = -0.13 \pm 0.02$ and
$\alpha^{8.4 GHz}_{5 GHz} = 0.39 \pm 0.04$ ($S \propto \nu^{-\alpha}$).
\par\noindent
The determination of the core and jet spectral index is difficult
in both frequency ranges, though for different reasons.
Between 1.7 GHz and 5 GHz we can
derive only the core spectral index, because of the very different u-v
coverage at these frequencies (see also Table 1 and Fig. 1). From the full
resolution maps we obtain $\alpha_{1.7 GHz}^{5 GHz}$
\ltsim 0.07 $\pm 0.09$. This value should be considered an
upper limit, since the 1.7 GHz core flux is likely to be contaminated
by the jet emission.
\par\noindent
In order to derive the core and jet spectral index between
5 GHz and 8.4 GHz we obtained maps (not shown here) using similar u-v
intervals and
convolved with the same restoring beam. This procedure,
however, leads to a rather poor u-v coverage at both frequencies, degrading the
quality of the two maps. Furthermore the spacing sampling in the u-v plane
is quite different at 5 GHz and 8.4 GHz (see Table 1 and Fig. 1), resulting
in partial loss of extended structure in the high frequency map.
The most reliable values we obtained are $\alpha^{8.4 GHz}_{5 GHz} = 0.27 \pm
0.08$ for the core
and $\alpha^{8.4 GHz}_{5 GHz} = 0.9 \pm 0.2$ for the jet. Given the
uncertainties mentioned, the jet spectral index should be considered an
upper limit, and in our discussion (Sect. 4) we will use the conservative
value $\alpha$ = 0.5.
\medskip
\noindent
{\it 3.2.3 The kiloparsec-scale properties}
\par\noindent
Eilek et al. (1984) discussed the large scale properties of 3C465.
Here we want to focus our attention on the jet properties as visible from our
map (Fig. 6), before the jet culminates in the bright spot, at
$\sim 30^{\prime\prime}$ from the core (Fig. 7).
The kiloparsec-scale jets are very asymmetric, the counterjet
being barely visible on our 5 GHz map.
\par\noindent
Slices taken perpendicular to the direction of propagation of the main jet
indicate that the jet remains very collimated, i.e. \ltsim $1.1^{\prime\prime}$
out to $\sim 32^{\prime\prime}$ ($\sim$ 13.4 kpc)
from the core. After a marginal expansion, recollimation occurs at
$\sim 20^{\prime\prime}$ from the core. In this region
the measured jet diameter is
$\sim 0.6^{\prime\prime}$ and the brightness has a secondary peak,
i.e. $\sim 3$ mJy/beam.
\par\noindent
The brightness along the jet in 3C465
suggests that there is no regime change in the jet dynamics with
increasing distance from the core.
This is confirmed also by the polarization map, which shows
that the polarized vectors remain perpendicular to the jet direction
all the way to the {\it bright spot}.
(Fig. 7).
\par\noindent
The jet/counterjet brightness ratio $R$ derived from our map
suggests that $R$ remains fairly constant, i.e. $R \sim$ 20, all along
the jet.
\vskip 0.8truecm
\noindent
{\bf 4. Discussion}
\medskip
\noindent
The two FR-I radio galaxies presented in this paper
exhibit similar global radio properties. They are both
characterised by straight radio jets, asymmetric in brightness, and by
tails of diffuse radio emission. The distortions of the radio morphology
take place along the tails, therefore for the following discussion
we will assume that the jets are straight on both parsec and kiloparsec scale.
\medskip
\par\noindent
{\it 4.1 The parsec-scale asymmetry}
\par\noindent
The sources B2 0836+29 and 3C465 are asymmetric on the parsec scale.
They are both
characterised by a core-jet morphology, with the milliarcsecond jet aligned
within few degrees with the main arcsecond scale one.
As in similar cases
(Giovannini et al., 1994), this is interpreted as due to Doppler favoritism
in intrinsically symmetric relativistic jets. Under this
assumption we will use the available data to derive constraints on the
plasma speed and on its orientation to the line of sight, and we will
compare the results to the predictions of the unified models.
\par\noindent
Under the hypothesis of intrinsic symmetry for the radio jets and
the observed asymmetry $R$ due to Doppler
favoritism, we can constrain the product
$\beta cos \theta$ ($\beta = {v \over c}$, and $\theta$ is the angle of
the source radio axis to the line of sight) by means of the
standard formula $\beta cos \theta = (R^{\xi} - 1)/(R^{\xi} + 1)$
(Pearson and Zensus, 1987), where  the parameter
$\xi$ is defined as $\xi = {1 \over {2+\alpha}}$, $\alpha$ being the spectral
index of the radio emission.
\par\noindent
Assuming an isotropic jet emissivity (Giovannini et al., 1994)
and a spectral index $\alpha$ = 0.5 for both sources,
the milliarcsecond asymmetry found for B2 0836+29 and 3C465
gives $\beta cos \theta~\gtsim~0.54$, and $\beta cos \theta~\gtsim~0.59$
respectively.
The derived allowed ranges for $\beta$ and $\theta$ are given in
Figs. 8a and 8b.
\par\noindent
A second independent contraint on the angle to the line of sight
and on the jet velocity is derived by means
of the ratio between the 5 GHz core power and the 408 MHz total power.
In order to derive an upper and lower limit for $\beta$ and $\theta$
we used the correlation (Giovannini et al., 1994):
$$\beta = (K-1)(K cos\theta -0.5 )^{-1}$$
with $K = [P_c(\theta)/P_c(60)]^{0.5}$.
Here $P_c$ is the measured (beamed) radio
power, and $P_c(60)$ is the apparent (beamed) radio power for a galaxy oriented
at 60$^{\circ}$ with respect to the line of sight. We have taken into account
the statistical uncertainties (1$\sigma$) and a possible core flux density
variability up to a factor of 2.
In both sources the core is prominent, suggesting a small angle to the line
of sight, as discussed below.
The results of this analysis are also reported in Figs. 8a and 8b, which
allow us to draw the following conclusions:
\par\noindent
- for B2 0836+29 the core prominence is high, and this
plays the main role, constraining
$\theta$ \ltsim $37^{\circ}$, and the jet velocity v$_{jet}$ \gtsim 0.75c;
\par\noindent
- for 3C465 the combination of the core prominence and the jet/counterjet
brightness ratio give $\theta$ \ltsim $54^{\circ}$, and
v$_{jet}$ \gtsim 0.6c.
\par\noindent
The parsec-scale morphology and the velocity derived for the plasma speed
in the vicinity of the radio nucleus for these two sources confirms the
fact that FR-I and FR-II radio galaxies are indistinguishable on the parsec
scale, both in their morphology and properties,
as already pointed out in Paper IV. They also confirm that jets start
out relativistically even in low-intermediate luminosity radio galaxies
and decelerate further out, as expected by current models
(Laing, 1993; Bicnkell, 1995).
\par\noindent
BL-Lac objects are characterised by intrinsic $\gamma$ factors
($\gamma = {1 \over {\sqrt{1-\beta^2}}}$) of the order of a few units
(Ghisellini et al., 1993).
If FR-I radio galaxies are
the parent population of BL-Lac objects, the range of their intrinsic
$\gamma$ should be the same. We point out that for the sources presented here
we have $\gamma$ \gtsim 3 if 25$^{\circ}$ \ltsim $\theta$ \ltsim 37$^{\circ}$
for B2 0836+29, and 35$^{\circ}$ \ltsim $\theta$ \ltsim 54$^{\circ}$ for
3C465. These numbers are in agreement with the expectations derived
from the unification models, which predict $\theta$ \gtsim $30^{\circ}$
(see Ghisellini et al., 1993).
\par\noindent
The analysis of the X-ray emission coming from the nuclear regions
and the application of the
synchrotron-self-Compton (SSC) model (Marscher 1987) is being done for
these two radio galaxies, as well as for the other galaxies in our sample
(Giovannini, Feretti \& Comoretto, 1990) using either ROSAT data or
ad-hoc observations. The results will be presented in a separate paper.
\par\noindent
No constraints can be derived from the global size of the two radio
sources, since they are both characterised by a distorted
morphology, possibly caused by the combination of the galaxy motion and
of the density of the intergalactic medium.
\medskip
\noindent
The high plasma speed derived on the parsec
scale makes these two sources good candidates for the detection
of motion of the components along the jets. In particular,
assuming that what we see on the parsec scale is the plasma flow,
the standard beaming model suggests that the apparent motion $\beta_{app}$
of the VLBI components is related to the intrinsic plasma speed
$\beta_{intr}$ by the formula
$$\beta_{app} = {{\beta_{intr} sin\theta} \over {1 - \beta_{intr} cos\theta}}$$
(Pearson \& Zensus, 1987).
If we assume for both sources an angle of $35^{\circ}$ to the line of sight,
the formula reported above gives
an apparent proper motion $\mu \sim 0.32$
mas/year for B2 0836+29 and $\sim$ 0.46 mas/yr for 3C465.
\par\noindent
In order to see if any motion is visible in the components of the parsec-scale
jet, a second epoch observation at 5 GHz and 8.4 GHz respectively for B2
0836+29
and 3C465 has already been done. The results of the comparison between the
first and second epoch for these two sources will be presented in the future.
\medskip
\noindent
{\it 4.2 The kiloparsec-scale asymmetry}
\par\noindent
We derived the constraints on $\theta$ and $\beta$ along the kiloparsec
scale jet under the same assumption of Doppler boosting.
The straight morphology of the kiloparsec-scale jets suggests that jet bending
is not prominent here, and justifies our approach.
\par\noindent
For B2 0836+29 the jet/counterjet brightness ratio derived from our map
gives a value of $\beta cos \theta$ decreasing with
increasing distance from the core, i.e. from $\beta cos \theta$ = 0.36 at
$\sim$ 15 kpc from the core, to $\beta cos \theta$ = 0.14  at $\sim$ 30 kpc
(see Tab. 4). Assuming that the viewing
angle $\theta$ does not change, as suggested by the source structure,
the trend in $\beta cos \theta$ implies a decrease in the jet velocity
at increasing distance from the core.
In 3C465, $\beta cos \theta$ remains constant, i.e. 0.54, all along the jet.
\par\noindent
If we assume that $\theta$ does not change going from the parsec to the
kiloparsec scale,
the derived jet velocity on the large scale for B2 0836+29 is
0.44c at 15 kpc from the core and decreases to 0.17c at 30 kpc.
These numbers  are in good agreement with those found by
CFP95, who derived $25^{\circ}$ \ltsim $\theta$ \ltsim $45^{\circ}$ at 5 kpc
from the core, with a velocity in the range 0.5c - 0.85c.
On the contrary, in 3C465 the jet velocity remains pretty constant, i.e. 0.7c,
all the way to the {\it bright spot} (13 kpc from the core).
The different properties of the jet velocity in the two sources, i.e.
decreasing
in B2 0836+29, and constant in 3C465, are in agreement with other properties
of the kiloparsec scale jet, such as for example the opening angle of the
jet and the direction of magnetic field vector, presented in
{\it Sects. 3.1.3} and {\it 3.2.3}.  In B2 0836+29 the
opening angle of the jet increases going further out along the jet,
and the magnetic
field switches from longitudinal to transverse. On the other hand, in 3C465
the jet remains very collimated all its length long, and no change in
the polarization vector is observed. We want to note here that a study
on a sample of 3CR quasars carried out by Bridle et al. (1994), shows
evidence of deceleration from the parsec to kiloparsec scale even in high power
radio jets.
\medskip
\noindent
{\it 4.3 The unified models}
\par\noindent
The similarity of the parsec-scale properties in B2 0836+29 and 3C465,
compared to the differences
found on the large scale, suggests that the central engine in these two
radio galaxies may be the same, but that differences in their interaction with
the surrounding medium could be responsible for the different morphology and
properties of the kiloparsec-scale jets.
\par\noindent
In Paper IV we have shown that FR-I and FR-II radio
galaxies are very similar on the parsec scale, despite their large differences
on the large scale.
The results presented in this paper, together with
other examples in the literature for other FR-Is (i.e. M87, Biretta 1993,
and NGC6251, Jones \& Wehrle, 1995) reinforce the growing evidence
that the central engine in FR-Is and FR-IIs must be
similar.
The different radio morphology could then be due either
to an intrinsically  different power and torus (Falcke et al., 1995), or
to different conditions in the region beyond the parsec scale, where a
significant deceleration in the FR-I jets occurs (De Young, 1993;
Bicknell, 1995).
We point out that Maraschi \& Rovetti (1994) showed that
BL-Lacs and flat spectrum radio quasars can be linked together in a single
beamed population, their parent unbeamed population including FR-Is, FR-IIs and
steep spectrum radio quasars.
\par\noindent
More radio
observations, at sub-arcsecond resolution (using for example
the MERLIN or an upgraded VLA) and at very high resolution
(with satellite VLBI) are necessary to
firmly discriminate between these two hypothesis.
\par\noindent
\vskip 0.8truecm
\noindent
{\bf 5. Conclusions}
\par\noindent
We can summarize the results presented in this paper as follows.
\par\noindent
1. The parsec-scale core-jet morphology found for B2 0836+29 and for 3C465
confirms that
this is the most common parsec-scale strucure found in low-intermediate
luminosity radio galaxies, and that this class of radio sources is
indistinguishable from the more luminous FRIIs on the small scale.
\par\noindent
2. The values derived for the plasma speed are $\beta$ \gtsim $0.75$ for
B2 0836+29 and $\beta$ \gtsim $0.6$ for 3C465.
The limits to the angle to to line of sight are
$\theta$ \ltsim $37^{\circ}$ and
\par\noindent
$\theta$ \ltsim $54^{\circ}$ respectively
for B2 0836+29 and 3C465. These angles are consistent with a Doppler
factor $\gamma$ \gtsim 3.
\par\noindent
3. The intrinsic velocities derived for the plasma speed are in agreement
with the expectations, both theoretical
and observational, that
jets start out at relativistic speed even in low-intermediate luminosity radio
galaxies and decelerate on the kiloparsec scale.
\par\noindent
4. Both the velocities and the angles to the line of sight derived in these two
sources support the idea that low luminosity radio galaxies are the unbeamed
parent population of BL Lac radio sources.
\par\noindent
5. Finally, the results presented in this paper support the idea that
the central
engine in FRI and FRII radio galaxies is essentially the same, and that
the differences in the large scale properties for the two classes of radio
sources could be related to an intrinsically different core power or/and
differences in the environment outside the nuclear region of the associated
optical galaxy.
\vskip 0.8truecm
\noindent
\centerline{\bf Acknowledgments}
\par\noindent
We thank the staffs at the telescopes for their assistance and the people
assisting us at the California Institute of Technology
and in Bonn, where the data were correlated absenteee by the local staff.
We acknowledge D. Dallacasa, F. Mantovani and R. Fanti for critical reading of
the manuscript.
Thanks are due to V. Albertazzi for drawing the figures and to N. Primavera
for taking the photographs.
T.V., J.M.M. and L.L. gratefully acknowledge the receipt of a
grant from CNR/CSIC (Prot. n. 127228).
\par\noindent
The National Radio Astronomy Observatory is operated by
Associated Universities, Inc., under contract with the National Science
Foundation.
\vfill
\eject
\noindent
{\bf References}
\medskip
\parindent -20pt
Alef, W., 1989, in Very Long Baseline Interferometry, ed. M. Felli \& R.E.
Spencer (Dordrect: Kluwer), p. 97
\goodbreak
\parindent -20pt
Antonucci, R.R.J., 1993, ARA\&A 31, 473
p. 145
\goodbreak
\parindent -20pt
Bicknell, G.V., 1995 ApJ, in press
\goodbreak
\parindent -20pt
Biretta, J.A., in Astrophysical Jets, ed. D. Burgarella, M. Livio \& C.P.
O'Dea (Cambridge University Press), 263
\goodbreak
\parindent -20pt
Bridle. A.H., Hough, D.H., Lonsdale, C.J., Burns, J.O., Laing, R.A., 1994,
The Astron. J., 108, 766
\goodbreak
\parindent -20pt
Burns, J.O., Roettinger, K., Pinkney, J., Loken., C., Doe, S., Owen, F.,
Voges, W., White., R., 1994, in The Soft X-Ray Cosmos, AIP Conf. Proc. 313
ed. E. Schlegel \& R. Petre, NY:AIP, p. 183
\goodbreak
\parindent -20pt
Capetti A., Morganti, R., Parma, P., Fanti, R., 1993, A\&ASS 99, 407
\goodbreak
\parindent -20pt
Capetti, A., Fanti, R., Parma, P., 1995, A\&A, in press
\goodbreak
\parindent -20pt
Cotton, W.D., Giovannini, G., Feretti, L., Venturi, T., Lara, L., Marcaide,
J.M., Wehrle, A.E., 1995, ApJ, in press, Paper V
\goodbreak
\parindent -20pt
De Robertis, M.M., Yee, H.K.C., 1990, AJ 100, 84
\goodbreak
\parindent -20pt
De Young, D.S., 1993, ApJ 405, L13
\goodbreak
\parindent -20pt
Dressler A., 1980, ApJS, 42,565
\goodbreak
\parindent -20pt
Ekers, R.D., Fanti, R., Miley, G.K., 1983, A\&A, 120, 297
\goodbreak
\parindent -20pt
Eilek, J.A., Burns, J.O., O'Dea, C.P., Owen, F.N., 1984, ApJ 278, 37
\goodbreak
\parindent -20pt
Falcke, H., et al, 1995 A\&A, in press
\goodbreak
\parindent -20pt
Fanaroff, B.L., Riley, J.M., 1974, MNRAS, 167, 31
\goodbreak
\parindent -20pt
Fanti, C., Fanti, R., de Ruiter, H.R., Parma, P., 1987, A\&AS 69, 57
\goodbreak
\parindent -20pt
Feretti, L., Giovannini, G., Venturi, T., Wehrle, A.E., 1993, ApJ 408, 446,
Paper III
\goodbreak
\parindent -20pt
Ghisellini, G., Padovani, P., Celotti, A., Maraschi, L., 1993, ApJ 407, 65
\goodbreak
\parindent -20pt
Giovannini, G., Feretti, L., Gregorini, L., Parma, P., 1988, A\&A 199, 73
\goodbreak
\parindent -20pt
Giovannini, G., Feretti, L., Comoretto, G., 1990, ApJ 358, 159, Paper I
\goodbreak
\parindent -20pt
Giovannini, G., Feretti, L., Venturi, T., Lara, L., Marcaide, J.M., Rioja,
M.J.,
Spangler, S.R., Wehrle, A.E, 1994, ApJ 435, 116, Paper IV
\goodbreak
\parindent -20pt
Jones, D.L., Wehrle, A.E., 1995, ApJ, in press
\goodbreak
\parindent -20pt
Laing, R.A., 1993, in Astrophysical Jets, Eds. Fall, M., O'Dea, C., Livio, M.,
Burgarella, D., p. 26
\goodbreak
\parindent -20pt
Maraschi, L., Rovetti, F., 1994, ApJ 436, 79
\goodbreak
\parindent -20pt
Marscher, A.P., 1987, in Superluminal Radio Sources, ed. J.A. Zensus \&
T.J. Pearson (Cambridge University Press), 280
\goodbreak
\parindent -20pt
Morganti, R., Fanti, R., Gioia, I.M., Harris, D.E., Parma, P., de Ruiter, H.,
1988, A\&A 189,11
\goodbreak
\parindent -20pt
O'Donoghue, A.A., Eilek, J.A., Owen, F.N., 1993, ApJ 408, 428
\goodbreak
\parindent -20pt
O'Donoghue, A.A., Owen, F.N., Eilek, J.A., 1990, ApJ Suppl. Ser. 72, 75
\goodbreak
\parindent -20pt
Owen, F.N., Rudnick, L., 1976, ApJ, 205, L1
\goodbreak
\parindent -20pt
Owen, F.N., Rudnick, L., Peterson, B.M., 1977, AJ 82, 677
\goodbreak
\parindent -20pt
Owen, F.N., White, R.L., Thronson, H.A., 1988, AJ 95, 1
\goodbreak
\parindent -20pt
Pearson, T.J., 1991, BAAS 23, 991
\goodbreak
\parindent -20pt
Pearson, T.J., Zensus, A.J., 1987, in Superluminal Radio Sources, eds. Pearson
\& Zensus (Cambridge University Press), p. 1
\goodbreak
\parindent -20pt
Schwab, F.R., Cotton, W.D., 1983, AJ 88, 688
\goodbreak
\parindent -20pt
van Breugel, W.J.M., 1980, A\&A 88, 248
\goodbreak
\parindent -20pt
Venturi, T., Giovannini, G., Feretti, L., Comoretto, G., Wehrle, A.E., 1993,
ApJ 408, 81, Paper II
\goodbreak
\parindent -20pt
Zhao, J.-H., Burns, J.O., Owen, F.N., 1989, AJ 98, 1
\vfill
\eject
\centerline{\bf Figure Captions}
\medskip
\noindent
{\bf Figure 1 -} uv-coverage of our VLBI data sets. a) B2 0836+29  at 5 GHz;
b) 3C465 at 1.7 GHz; c) 3C465 at 5 GHz; d) 3C465 at 8.4 GHz.
\medskip
\noindent
{\bf Figure 2 -} Contour plot of B2 0836+29  on the VLBI scale.
The peak flux in the map is 128 mJy/beam.
Contour levels are -0.4, 0.4, 0.6, 0.8,
1, 1.5, 2, 3, 4, 5, 7, 10, 20, 30, 50, 70, 100, 120 mJy/beam. The restoring
beam is $1.9\times0.7$ mas in p.a. $0^{\circ}$.
\medskip
\noindent
{\bf Figure 3 -} VLA C array total intensity map of B2 0836+29 at 5 GHz.
The peak flux
is 145 mJy/beam. Contour levels are -0.2, -0.1, 0.1, 0.2, 0.3, 0.5, 1,
2, 5, 10, 20, 30, 50, 100, 120 mJy/beam. The restoring beam is
$3.67^{\prime\prime} \times 3.58^{\prime\prime}$ in p.a. $-84.5^{\circ}$.
The noise in the map is 0.03 mJy/beam.
\medskip
\noindent
{\bf Figure 4 -} Percentage of polarization for B2 0836+29, superimposed to
the total intensity map. 1 arcsec corresponds to
5$\times10^{-5}$ Jy/beam.
The total intensity peak is
145 mJy/beam. Contour levels are -0.1, 0.1, 0.5, 1,
5, 30, 100 mJy/beam. The restoring beam is
$3.67^{\prime\prime} \times 3.58^{\prime\prime}$ in p.a. $-84.5^{\circ}$.
\medskip
\noindent
{\bf Figure 5 -} Contour plot of 3C465 on the VLBI scale.
{\bf a)} 1.7 GHz map. The peak flux in the map is 165 mJy/beam. Contour levels
are -1, 1, 1.5, 2, 3, 5, 10, 20, 30, 50, 100 mJy/beam. The restoring
beam is $12.5\times11.2$ mas in p.a. $85^{\circ}$. The noise in the map
is 0.45 mJy/beam.
{\bf b)} 5 GHz map. The peak flux in the map is 133 mJy/beam. Contour levels
are
-0.75, 0.75, 1.5, 2, 3, 5, 10, 20, 30, 50, 100 mJy/beam. The restoring
beam is $2.52\times0.83$ mas in p.a. $-9.7^{\circ}$. The noise in the
map is 0.33 mJy/beam.
{\bf c)} 8.4 GHz map. The peak flux in the map is 132 mJy/beam.
Contour levels are
-0.75, 0.75, 1.5, 2, 3, 5, 10, 20, 30, 50, 100 mJy/beam. The restoring
beam is $2.52\times0.83$ mas in p.a. $-9.7^{\circ}$. The noise in the
map is 0.35 mJy/beam.
\medskip
\noindent
{\bf Figure 6 -} 5 GHz VLA C array total intensity map of 3C465.
The peak flux
is 242 mJy/beam. Contour levels are -0.2, 0.2, 0.5,  1,
2, 3, 4, 6, 10, 15, 20, 30, 50, 100, 200
mJy/beam. The restoring beam is
$3.87^{\prime\prime} \times 3.54^{\prime\prime}$ in p.a. $-80.7^{\circ}$.
The noise in the map is 0.05 mJy/beam.
\medskip
\noindent
{\bf Figure 7 -} Percentage of polarization for 3C465 superimposed to
the total intensity map.
1 arcsec corresponds to $3.3\times10^{-4}$ Jy/beam. The total intensity peak is
242 mJy/beam. Contour levels are 0.1, 0.3, 0.5, 1, 3, 5, 10, 50,
100, 200 mJy/beam. The restoring beam is
$3.87^{\prime\prime} \times 3.54^{\prime\prime}$ in p.a. $-80.7^{\circ}$.
\medskip
\noindent
{\bf Figure 8 -} Constraints on the angle $\theta$ and the intrinsic jet
velocity $\beta$ for {\bf a)} B2 0836+29 and {\bf b)} 3C465.
A and A$^{\prime}$ respectively the lower and upper limits derived from the
ratio between the core flux density and the total flux density. B
is the limit derived from the jet/counterjet brightness ratio.
The allowed region is the undashed one.
%
%
\nopagenumbers
\parindent 0pt
\def\ltsim{\raise 2pt \hbox {$<$} \kern-1.1em \lower 4pt \hbox {$\sim$}}
\def\ltapprox{\raise 2pt \hbox {$<$} \kern-1.1em \lower 5pt \hbox {$\approx$}}
\def\gtsim{\raise 2pt \hbox {$>$} \kern-1.1em \lower 4pt \hbox {$\sim$}}
\def\gtapprox{\raise 2pt \hbox {$>$} \kern-1.1em \lower 5pt \hbox {$\approx$}}
\vskip 1.5truecm
\parindent 0pt
{\bf Table 1. Observational Parameters}
\vskip 0.5truecm
\noindent
$$\vbox {\settabs 7\columns \baselineskip 15pt
\hrule
\medskip
\+ \hfil (1) \hfil & \hfil (2) \hfil & \hfil (3) \hfil & \hfil (4) \hfil
& \hfil (5) \hfil  & \hfil (6) \hfil & \hfil (7) \hfil &\cr
\+ \hfil Source \hfil & \hfil $\nu$ \hfil & \hfil Mode \hfil & \hfil
Array \hfil & \hfil uv-range \hfil &
\hfil Obs. Time  \hfil & \hfil Epoch \hfil & \hfil &\cr
\+ &\hfil MHz \hfil &&& \hfil M$\lambda$ \hfil & \hfil hr \hfil &&\cr
\medskip
\hrule
\medskip
\+ \hfil 0836+290 \hfil & \hfil 4990.99 \hfil & \hfil MKIIIB \hfil &
\hfil EU-US$^{(a)}$ \hfil & \hfil 5-150 \hfil & \hfil 12 \hfil & \hfil
Nov. 90 \hfil &\cr
\+ \hfil 3C465 \hfil & \hfil 1662.99 \hfil & \hfil MKII \hfil &
\hfil Y,VLBA$^{(b)}$ \hfil & \hfil 0.2-12 \hfil & \hfil 12 \hfil & \hfil
Jan. 92 \hfil &\cr
\+ &\hfil 4990.99 \hfil & \hfil MKIIIB \hfil &
\hfil EU-US$^{(c)}$ \hfil & \hfil 4-150 \hfil & \hfil 10 \hfil & \hfil Mar. 92
\hfil &\cr
\+ &\hfil 8416.99 \hfil & \hfil MKII \hfil & \hfil LNM,Y,VLBA$^{(c)}$
\hfil & \hfil 1-260 \hfil & \hfil 12 \hfil & \hfil Jan. 92 \hfil &\cr
\medskip
\hrule
\medskip
}$$
\noindent
$^{(a)}$ EU = Bonn, WSRT, Onsala, Jodrell Bank, Medicina
\par
\hskip 0.6truecm
US = Green Bank, Owens Valley, VLA Phased Array
\par\noindent
$^{(b)}$ VLBA = KP,PT,LA,FD,NL,BR,OV
\par\noindent
$^{(c)}$ EU = Bonn, WSRT, Jodrell Bank, Medicina, Noto
\par
\hskip 0.6truecm
US = Haystack, Green Bank, Pie Town, North Liberty, VLA Phased Array
\vskip 1.0truecm
\parindent 0pt
{\bf Table 2. VLBI Flux Densities}
\vskip 0.5truecm
\noindent
$$\vbox {\settabs 6\columns \baselineskip 15pt
\hrule
\medskip
\+ \hfil (1) \hfil & \hfil (2) \hfil & \hfil (3) \hfil & \hfil (4) \hfil
& \hfil (5) \hfil & \hfil (6) \hfil &\cr
\+ \hfil Source \hfil & \hfil $\nu$  \hfil & \hfil Beam \hfil & \hfil Noise
\hfil & \hfil S$_{core}$ \hfil & \hfil S$_{total}$ \hfil &\cr
\+ &\hfil MHz \hfil & \hfil mas, (${\circ}$) \hfil & \hfil mJy/b \hfil &
\hfil mJy \hfil & \hfil mJy \hfil &\cr
\medskip
\hrule
\medskip
\+ \hfil 0836+290 \hfil & \hfil 4990.99 \hfil & \hfil 1.9$\times$0.7, (0)
\hfil & \hfil  0.18 \hfil & \hfil 146 $\pm$ 7 \hfil & \hfil 167 $\pm$ 8
\hfil &\cr
\+ \hfil 3C465 \hfil & \hfil 1662.99 \hfil & \hfil 12.5$\times$11.1,$~(85)$
\hfil & \hfil 0.45 \hfil & \hfil  181 $\pm$ 9
\hfil &  \hfil 203 $\pm$ 10 \hfil &\cr
\+ &\hfil 4990.99 \hfil & \hfil 2.52$\times$0.83,$~(-9.7)$ \hfil & \hfil 0.33
\hfil & \hfil 168 $\pm$ 7 \hfil &
\hfil 234 $\pm$ 12 \hfil &\cr
\+ &\hfil 8416.99 \hfil & \hfil 2.52$\times$0.83,$(~-9.7)$ \hfil & \hfil 0.35
\hfil  & \hfil 146 $\pm$ 7 \hfil &
\hfil 191 $\pm$ 10 \hfil &\cr
\medskip
\hrule
\medskip
}$$
\vskip 1.0truecm
\noindent
{\bf Table 3. VLA Data}
\vskip 0.5truecm
\noindent
$$\vbox {\settabs 7\columns \baselineskip 15pt
\hrule
\medskip
\+ \hfil (1) \hfil & \hfil (2) \hfil & \hfil (3) \hfil & \hfil (4) \hfil
& \hfil (5) \hfil & \hfil (6) \hfil & \hfil (7) \hfil &\cr
\+ \hfil Source  \hfil & \hfil RA$_{core}$ \hfil & \hfil DEC$_{core}$
\hfil & \hfil $\nu$ \hfil & \hfil Beam \hfil & \hfil
rms \hfil & \hfil S$_{core}$ \hfil &\cr
\+ &\hfil (J2000) \hfil & \hfil (J2000) \hfil  &\hfil MHz \hfil & \hfil
$^{\prime\prime}$,$^{\circ}$ \hfil & \hfil mJy/b \hfil & \hfil mJy \hfil &\cr
\medskip
\hrule
\medskip
\+ \hfil 0836+290 \hfil & \hfil $08^h39^m15.827^s$ \hfil & \hfil
$28^{\circ}50^{\prime}38.86^{\prime\prime}$ \hfil & \hfil
4990.99 \hfil & \hfil 3.67$\times$3.58, -84.5
\hfil & \hfil 0.025 \hfil & \hfil 161 $\pm$ 8 \hfil &\cr
\+ \hfil 3C465 \hfil & \hfil $23^h38^m29.376^s$ \hfil & \hfil
$27^{\circ}01^{\prime}53.31^{\prime\prime}$ \hfil & \hfil
4990.00 \hfil & \hfil 3.87$\times$3.54, -80.7
\hfil & \hfil 0.05 \hfil & \hfil 246 $\pm$ 12 \hfil &\cr
\medskip
\hrule
\medskip
}$$
\vfill
\eject
\noindent
{\bf Table 4. Jet Constraints}
\vskip 0.5truecm
\noindent
$$\vbox {\settabs 6\columns \baselineskip 15pt
\hrule
\medskip
\+ &\hfil $~~~~~~~~~~~~~~~~$ VLBI \hfil && \hfil $~~~~~~~~~~~~~~~~~~~~~~~~~~~~$
VLA \hfil &&\cr
\+ &\hfil $\beta cos \theta$ \hfil & \hfil d (pc)  \hfil &
\hfil $\beta cos \theta$ \hfil & \hfil d ($^{\prime\prime}$) \hfil
& \hfil D (kpc) \hfil &\cr
\medskip
\hrule
\medskip
\+ \hfil B2 0836+29 \hfil & \hfil \gtsim 0.54 \hfil & \hfil 6 \hfil & \hfil
0.36
\hfil & \hfil $\le$ 15 \hfil & \hfil $\le$ 15 \hfil &\cr
\+ &&&\hfil 0.27 \hfil & \hfil 15-20  \hfil & \hfil 15-20 \hfil &\cr
\+ &&&\hfil 0.14 \hfil & \hfil 20-31  \hfil & \hfil 20-31 \hfil &\cr
\+ \hfil 3C465 \hfil & \hfil \gtsim 0.59 \hfil & \hfil 1 \hfil & \hfil
0.54  \hfil & \hfil $\le$ 30 \hfil & \hfil $\le$ 13 \hfil &\cr
\medskip
\hrule
\medskip
}$$
\bye